\documentclass[cits]{PoS}
\pdfoutput=1
\usepackage{amsfonts}
\usepackage[centertags]{amsmath}
\usepackage{amssymb,amscd,amsbsy}
\usepackage{eurosym}
\usepackage{./styles/mcite}
\usepackage{./styles/myxspace}
\usepackage{./styles/mysidecaption}
\def\zero{{\scriptscriptstyle 0}}

\catcode`\@=11 
\@ifundefined{euro}%
{
 }%
{
 }
\catcode`\@=12 

\def\Z0{{Z^\zero}}
\def\eVdist{\kern-0.06667em}

\def\hour{\,\text{h}}

\def\IP{{\rm I$\kern-0.01667em$P}\xspace}

\def\Ptlj{{\not{\kern-0.55ex P}}_t\ell j}
\def\Ptmiss{{\not{\kern-0.55ex P}}_t}

\mathchardef\qsm=63
\mathchardef\pls=43
\mathchardef\mns=512
\mathchardef\plm=518
\mathchardef\eql=61
\mathchardef\smallleft=300
\mathchardef\smallright=301
\mathchardef\perslsh=47
\mathchardef\les=316
\mathchardef\gre=318
\mathchardef\leq=532
\mathchardef\grq=533
\chardef\usc=95
\chardef\til=126
\chardef\amp=38

\def\sqr#1#2#3{{\vcenter{\hrule height.#3ex\hbox{\vrule width.#2ex height#1ex
    \kern#1ex\vrule width.#3ex}\hrule height.#2ex}}}

\def\angleto{\vrule width.035em height2.1ex depth-.56ex\unskip\kern-.6ex\to}
\def\perchc#1{{\raise.4ex\hbox{$\mkern4mu#1{\it\perslsh}_
             {\mkern-5mu\scriptscriptstyle{{\rm o}\!{\rm o}}}^
             {\mkern-12.8mu\scriptscriptstyle{\rm o}}$}}}

\def\widebar#1{\mkern1.5mu\overline{\mkern-1.5mu#1\mkern-1.mu}\mkern1.mu}
\catcode`\@=11 
\def\parenbar{\mathpalette\p@renb@r}
\def\p@renb@r#1#2{\vbox{%
  \ifx#1\scriptscriptstyle \dimen@.7em\dimen@ii.2em\else
  \ifx#1\scriptstyle \dimen@.8em\dimen@ii.25em\else
  \dimen@1em\dimen@ii.4em\fi\fi \offinterlineskip
  \ialign{\hfill##\hfill\cr
    \vbox{\hrule width\dimen@ii}\cr
    \noalign{\vskip-.3ex}%
    \hbox to\dimen@{$\mathchar300\hfil\mathchar301$}\cr
    \noalign{\vskip-.3ex}%
    $#1#2$\cr}}}
\catcode`\@=12 
\def\nuan{\parenbar{\nu}\kern-0.4ex}

\def\nubar{\widebar{\nu}}

\newbox\struttbox
\setbox\struttbox=\hbox{\vrule height1.65ex depth.485ex width0pt}
\def\strutt{\relax\ifmmode\copy\struttbox\else\unhcopy\struttbox\fi}
\def\stru#1#2{\relax\ifmmode\hbox{\vrule height#1 depth#2 width0pt}
\else\vrule height#1 depth#2 width0pt\fi}
\def\uline#1{$\underline{\hbox{#1\strutt}}$}
\def\ronum#1{\uppercase\expandafter{\romannumeral#1}}
\def\ronuml#1{\expandafter{\romannumeral#1}}

\def\cbk{\kern-0.5em}

\newcommand{\linebox}[2][3.ex]{\uline{\hbox to #2{\stru{#1}{0.pt}\hfil}}}

\newcounter{seqnum}
\DeclareMathAlphabet{\mathbf}{OT1}{cmr}{bx}{n}

\DeclareMathAlphabet{\mathbfs}{OT1}{lcmss}{bx}{sl}

\newcommand{\PreserveBackslash}[1]{\let\temp=\\#1\let\\=\temp}
\newlength\listtextwidth

\catcode`\@=11 
\newlength{\@tabfninsert}
\newlength{\@tabfnwidth}
\newcommand{\tabfootnote}[2]{%
  \setlength{\@tabfninsert}{0.8em}
  \setlength{\@tabfnwidth}{\textwidth}
  \addtolength{\@tabfnwidth}{-\@tabfninsert}
  \addtolength{\@tabfnwidth}{-0.4em}
  \noindent\makebox[\@tabfninsert][r]{\footnotesize$^{#1}$\hfil}\hfill%
  \parbox[t]{\@tabfnwidth}{\footnotesize #2\hfill}}
\catcode`\@=12 
\newcommand{\boldarrayrulewidth}{1pt}

\catcode`\@=11 
\let\tab@penalty\relax
\newcount\tab@state
\def\bcline#1{%
  \noalign{\kern-.5\arrayrulewidth\tab@penalty}%
  \omit%
  \global\tab@state\@ne%
  \ranges\bcline@i{#1}%
  \cr%
  \noalign{\kern-.5\arrayrulewidth\tab@penalty}%
}
\def\bcline@i#1#2{%
  \ifnum#1<\tab@state\relax%
    \tab@@cr%
    \noalign{\kern-\arrayrulewidth\tab@penalty}%
    \omit%
    \global\tab@state\@ne%
  \fi%
  \@whilenum\tab@state<#1\do{%
    \hfil\tab@@tab@omit%
    \global\advance\tab@state\@ne%
  }%
  \ifnum\tab@state>\@ne%
    \kern-\arrayrulewidth%
  \fi%
  \@whilenum\tab@state<#2\do{%
    \tab@@span@omit%
    \global\advance\tab@state\@ne%
  }%
  \leaders\hrule\@height\boldarrayrulewidth\hfill%
}
\def\ranges#1#2{%
  \gdef\ranges@temp{#1}%
  \begingroup%
  \ranges@i#2 \q@delim%
}
\def\ranges@i{%
  \@ifnextchar\q@delim\ranges@done{\afterassignment\ranges@ii\count@}%
}
\def\ranges@ii{%
  \@ifnextchar-\ranges@iii{\ranges@do\count@\count@\ranges@v}%
}
\def\ranges@iii-{\afterassignment\ranges@iv\@tempcnta}
\def\ranges@iv{\ranges@do\count@\@tempcnta\ranges@v}
\def\ranges@v{%
  \@ifnextchar,%
    \ranges@vi%
    {%
      \@ifnextchar\q@delim%
        \ranges@done%
        {\tab@err@range\ranges@vi,}%
    }%
}
\def\ranges@vi,{\afterassignment\ranges@ii\count@}
\def\ranges@do#1#2{%
  \ifnum#1>#2\else%
    \expandafter\endgroup%
    \expandafter\ranges@temp%
    \expandafter{%
    \the\expandafter#1%
    \expandafter}%
    \expandafter{%
    \the#2%
    }%
    \begingroup%
  \fi%
}
\def\ranges@done\q@delim{\endgroup}
\def\ifinrange#1#2{%
  \@tempswafalse%
  \count@#1%
  \ranges\ifinrange@i{#2}%
  \if@tempswa%
    \expandafter\@firstoftwo%
  \else%
    \expandafter\@secondoftwo%
  \fi%
}
\def\ifinrange@i#1#2{%
  \ifnum\count@<#1 \else\ifnum\count@>#2 \else\@tempswatrue\fi\fi%
}
\def\tab@@cr{\cr}
\def\tab@@tab@omit{&\omit}
\def\tab@@span@omit{\span\omit}
\def\tab@checkrule#1{%
  \count@#1\relax%
  \expandafter\ifinrange%
  \expandafter\count@%
  \expandafter{\tab@xcols}%
    {\tab@checkrule@i}%
    {}%
}
\def\bhline{\noalign{\ifnum0=`}\fi\hrule \@height  
\boldarrayrulewidth \futurelet \@tempa\@xhline}
\def\@xhline{\ifx\@tempa\hline\vskip \doublerulesep\fi
      \ifnum0=`{\fi}}
\catcode`\@=12 

\catcode`\@=11 
\newcounter{pict@width}
\newcounter{pict@height}
\newlength{\pict@scale}
\newcommand{\psfigadd}[4]{%
\setcounter{pict@width}{1*\ratio{#2+\pict@scale/2}{\pict@scale}}
\setcounter{pict@height}{1*\ratio{#3+\pict@scale/2}{\pict@scale}}
\setlength{\unitlength}{\pict@scale}
\hbox to #2{\hspace{-\fill}\begin{picture}(\thepict@width,\thepict@height)
\put(0,0){\psfig{figure=#1,width=#2,height=#3,clip=}}
\SetScale{0.283466457}
\SetWidth{1.763889}
{#4}
\end{picture}}
}
\newcounter{pict@widthfst}
\newcounter{pict@widthscd}
\newcounter{pict@widthtot}
\newcommand{\psfigaddtwo}[7]{%
\setcounter{pict@widthfst}{1*\ratio{#2+\pict@scale/2}{\pict@scale}}
\setcounter{pict@widthscd}{1*\ratio{#2+#4+\pict@scale/2}{\pict@scale}}
\setcounter{pict@widthtot}{1*\ratio{#2+#4+#6+\pict@scale/2}{\pict@scale}}
\setcounter{pict@height}{1*\ratio{#3+\pict@scale/2}{\pict@scale}}
\setlength{\unitlength}{\pict@scale}
\hbox{\hspace{-\fill}\begin{picture}(\thepict@widthtot,\thepict@height)
\put(0,0){\psfig{figure=#1,width=#2,height=#3,clip=}}
\put(\thepict@widthscd,0){\psfig{figure=#5,width=#6,height=#3,clip=}}
\SetScale{0.283466457}
\SetWidth{1.763889}
{#7}
\end{picture}}
}
\newcommand{\psfigror}[4]{%
\setcounter{pict@width}{1*\ratio{#2+\pict@scale/2}{\pict@scale}}
\setcounter{pict@height}{1*\ratio{#3+\pict@scale/2}{\pict@scale}}
\setlength{\unitlength}{\pict@scale}
\hbox{\begin{picture}(\thepict@width,\thepict@height)
\put(0,\thepict@height){\psfig{figure=#1,width=#3,height=#2,clip=,angle=270}}
\SetScale{0.283466457}
\SetWidth{1.763889}
{#4}
\end{picture}}
}
\newcommand{\psfigrol}[4]{%
\setcounter{pict@width}{1*\ratio{#2+\pict@scale/2}{\pict@scale}}
\setcounter{pict@height}{1*\ratio{#3+\pict@scale/2}{\pict@scale}}
\setlength{\unitlength}{\pict@scale}
\hbox{\begin{picture}(\thepict@width,\thepict@height)
\put(0,0){\psfig{figure=#1,width=#3,height=#2,clip=,angle=90}}
\SetScale{0.283466457}
\SetWidth{1.763889}
{#4}
\end{picture}}
}
\catcode`\@=12 

\usepackage{graphicx}
\title{The ORCA Option for KM3NeT}
\ShortTitle{The ORCA Option for KM3NeT}
\author{\speaker{Ulrich F.\,Katz for the KM3NeT Collaboration}\thanks{www.km3net.org}\\
        Friedrich-Alexander University of Erlangen-Nürnberg,\\ 
        Erlangen Centre for Astroparticle Physics\\
        E-mail: \email{katz@physik.uni-erlangen.de}}
\abstract{
It has recently been suggested that the neutrino mass hierarchy can be
experimentally determined from the oscillation pattern of atmospheric neutrinos
passing through the Earth by measuring the two-dimensional arrival pattern of
neutrinos in energy and zenith angle, in the energy regime of about 3--20\,GeV. 
ORCA (Oscillation Research with Cosmics in the Abyss) is a study addressing the
feasibility of such a measurement employing the deep-sea neutrino telescope
technology developed for the KM3NeT project. In the following, the underlying
physics and resulting experimental signatures will be discussed and some aspects
of the ongoing simulation studies presented. A preliminary sensitivity estimate
derived from a simplified study strongly indicates that an exposure of at least
20\,Mton-years will be required to arrive at conclusive results. 
}
\FullConference{XV Workshop on Neutrino Telescopes,\\
		11-15 March 2013\\
		Venice, Italy}
\begin{document}
\section{Neutrino oscillations and mass hierarchy}
\label{sec:osc}

All current experimental information on neutrino oscillations can be described
correctly within a theoretical framework containing three neutrino types which
have different sets of flavour and mass eigenstates and different, hence
non-zero masses. The flavour eigenstates, denoted $\nu_\alpha$ with
$\alpha=e,\mu,\tau$, are relevant for weak interactions, i.e.\ neutrino
generation and reactions. The mass eigenstates ($\nu_i$ with masses $m_i$,
$i=1,2,3$) govern neutrino propagation through space-time.

The relation between flavour and mass eigenstates is given by the unitary
Pontecorvo-Maki-Nakagawa-Sakata matrix $U_{\alpha i}$, 
\begin{equation}
  |\nu_\alpha\rangle = \sum_{i=1,2,3}U_{\alpha i}|\nu_i\rangle\,.
  \label{eq:pmnsdef}
\end{equation}
A commonly used parameterisation of $U$ in terms of three mixing angles
$\theta_{12},\theta_{13},\theta_{23}$ and one complex phase $e^{i\delta}$ is
given by
\begin{equation}
  U=\begin{pmatrix}1&0&0\\0&c_{23}&s_{23}\\0&-s_{23}&c_{23}\end{pmatrix}\cdot
    \begin{pmatrix}c_{13}&0&s_{13}e^{-i\delta}\\0&1&0\\-s_{13}e^{i\delta}&0&c_{13}\end{pmatrix}\cdot
    \begin{pmatrix}c_{12}&s_{12}&0\\-s_{12}&c_{12}&0\\0&0&1\end{pmatrix}
  \label{eq:pmnspar}
\end{equation}
with $s_{ij}=\sin\theta_{ij}$ and $c_{ij}=\cos\theta_{ij}$. Two additional
complex phases appear in eq.~(\ref{eq:pmnspar}) if neutrinos are Majorana
particles. Since the Dirac or Majorana nature of neutrinos is unknown and these
phases are of no significance for the following discussion, they are omitted
here.

If the mixing angles and the mass-squared differences $\Delta
m^2_{ij}=m_i^2-m_j^2$ are non-zero, neutrino oscillations are a necessary
consequence. For propagation along a path length $L$ through vacuum the
probability $P_{\alpha\to\beta}$ for a flavour state $\alpha$ to turn into
flavour state $\beta$ can be calculated analytically from
\begin{equation}
  P_{\alpha\to\beta}=e^{ipL}\sum_{i=1,2,3}U_{\beta i}e^{-iE_it}U^\dagger_{i\alpha}\,
  \label{eq:oscvac}
\end{equation}
where $t$ the propagation time, $p$ the momentum and $E_i=\sqrt{p^2+m_i^2}$ the
energy of mass eigenstate $i$. Since the neutrino mass is experimentally
constrained to $m_i\lesssim2\,\text{eV}$, we can assume $p\gg m_i$ in the
following, hence in particular $t=L/c$ and $E_i=p+m_i^2/(2p)$. The complex phase
$e^{i\delta}$ introduces a CP-violating difference between oscillation
probabilities of neutrinos and antineutrinos. Neglecting these and other
sub-leading terms\footnote{Note that the identification of sub-leading terms
requires the knowledge of the mixing angles (see below).}, the resulting
$e\leftrightarrow\mu$ oscillation probability can be written as
\begin{equation}
  P_{e\to\mu}=P_{\mu\to e}=
  \sin^2\theta_{23}\sin^2(2\theta_{13})\sin^2\left(\frac{\Delta_{13}L}2\right)
  \label{eq:pemu}
\end{equation}
with $\Delta_{13}=\Delta m^2_{13}/(2E_\nu)$. A characteristic feature of vacuum
oscillations is that they do not depend on the signs of $\Delta m^2_{ij}$, i.e.\
they do not provide information on the mass ordering of the $\nu_i$.

For propagation in matter, e.g.\ through Sun or Earth, the oscillation pattern
is modified by the fact that $\nu_e$ have elastic interaction modes with
electrons that are not possible for other flavour eigenstates: In addition to
the $t$-channel exchange of a $Z$ boson (possible for all neutrino flavours),
elastic $\nu_ee$ ($\nubar_ee$) scattering can proceed through $u$ ($s$) channel
exchange of a $W$ boson. As a result, the forward scattering amplitude and hence
the index of refraction for $\nu_e$ is different from the other flavours. 
This effect is described quantitatively by an extra term in the Hamiltonian of
the $\nuan_e$,
\begin{equation}
  A=\pm\sqrt2G_FN_e\quad\text{with}\quad
    \begin{cases}+&\text{for $\nu_e$}\\-&\text{for $\nubar_e$}\,,\end{cases}
  \label{eq:adef}
\end{equation}
where $G_F$ is the Fermi constant and $N_e$ the number density of electrons in the
matter. The presence of this term modifies eq.~(\ref{eq:pemu}) to
\begin{eqnarray}
 &P^M_{e\to\mu}=P^M_{\mu\to e}=
    \sin^2\theta_{23}\sin^2(2\theta_{13}^\text{eff})
    \sin^2\left(\frac{\Delta^\text{eff}_{13}L}2\right)
  \label{eq:pemumat}\\
 &\sin^2(2\theta_{13}^\text{eff})=\sin^2(2\theta_{13})
    \cdot\frac{\Delta^2_{13}}{(\Delta^\text{eff}_{13})^2}
  \label{eq:theeff}\\
 &\Delta^\text{eff}_{13}=\sqrt{\left[\Delta_{13}\cos(2\theta_{13})-A\right]^2+
     \Delta^2_{13}\sin^2(2\theta_{13})}\,.
  \label{eq:deleff}
\end{eqnarray}
Note that via eq.~(\ref{eq:deleff}) the oscillation pattern now depends on the
sign of $\Delta_{13}$ and is different for neutrinos and antineutrinos; in
particular, we have 
\begin{equation}
  P^M_{e\to\mu}(\nu,\pm\Delta m^2_{13})=P^M_{e\to\mu}(\nubar,\mp\Delta m^2_{13})\,.
  \label{eq:pmatnuannu}
\end{equation}

\begin{figure}[hbt]
  \sidecaption
  \includegraphics[width=0.45\textwidth]{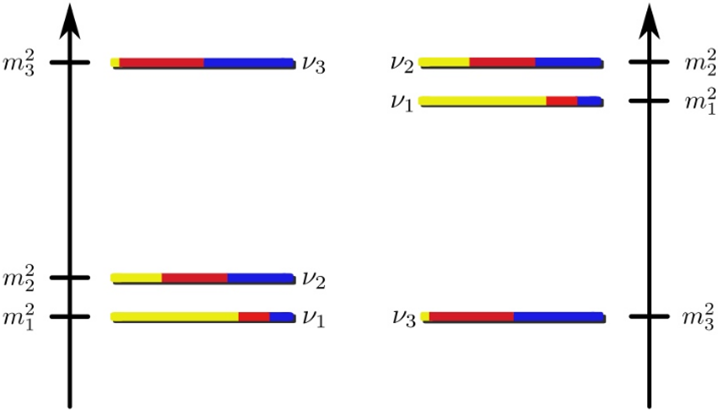}
  \caption{Graphic representation of neutrino masses and mixing, for the normal
           hierarchy (left) and the inverted hierarchy (right). The yellow (red, blue) 
           fractions of the lines indicate the $\nu_e$ ($\nu_\mu$, $\nu_\tau$) contents 
           of the mass eigenstates.}
  \label{fig:hierarchy}
\end{figure}

From measurements of atmospheric, solar, reactor and accelerator neutrinos the
parameters of neutrino oscillations are meanwhile rather precisely known (for
details see \cite{PDG-neutrinos-2013}). The mixing angles are roughly given by
$\sin\theta_{23}=\pi/4$, $\sin\theta_{12}=\pi/5.4$ and $\sin\theta_{13}=\pi/20$,
where the latter was the last to be measured about two years ago. The mass
differences are about $\Delta m^2_{12}=7.5\times10^{-5}\,\text{eV}^2$ and $|\Delta
m^2_{23}|=2.4\times10^{-3}\,\text{eV}^2$ (the third mass difference is not an
independent quantity). The sign of $\Delta m^2_{12}$ is known due to matter effects
in neutrino propagation through the Sun. The sign of $\Delta m^2_{23}$, however, is
as yet unknown, leaving two scenarios for neutrino mass ordering, denoted by
{\it normal hierarchy (NH)} and {\it inverted hierarchy (IH)}, respectively
(see Fig.~\ref{fig:hierarchy}).

A measurement of the neutrino mass hierarchy -- as one of the fundamental
parameters of the Standard Model of particle physics -- is an important goal in
itself. Beyond that, it would be important for easing the experimental
determination of the CP-violating phase $e^{i\delta}$ in future experiments, and
it could help in interpreting cosmological data and their dependence on the
neutrino sector.

\section{Measuring the neutrino mass hierarchy with atmospheric neutrinos}
\label{sec:atm}

Atmospheric neutrinos are mostly generated in $\pi$, $K$ and $\mu$ decays in
extended air showers initiated by cosmic ray interactions with nuclei in the
Earth atmosphere. In the atmospheric neutrino flux \cite{Gaisser-2002},
$\nu_\mu$ and $\nubar_\mu$ dominate over $\nu_e$ and $\nubar_e$, and at energies
beyond a few GeV muon neutrinos are slightly more abundant than muon antineutrinos.

\begin{figure}[hbt]
  \sidecaption
  \includegraphics[width=0.4\textwidth]{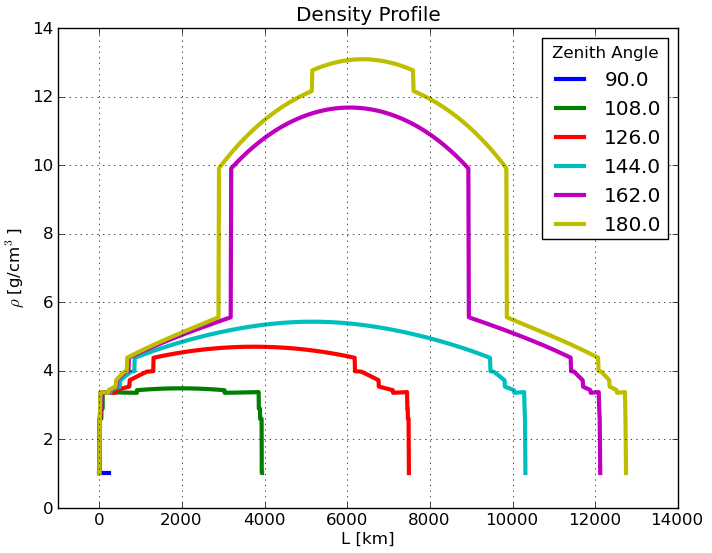}
  \caption{Density profiles of the Earth as traversed by neutrinos entering the 
           detector under different zenith angles. Zenith angles of $180^\circ$ 
           ($90^\circ$) correspond to vertically upward (horizontal) neutrino
           directions. The density data are taken from the {\it Preliminary Reference
           Earth Model (PREM)} \protect\cite{PREM-1981}.}
  \label{fig:prem}
\end{figure}

In order to exploit a matter-induced oscillation effect to distinguish NH and
IH, $\nuan_e$ must be involved. This is in particular the case for
$\nu_\mu\leftrightarrow\nu_e$ transitions which can be assessed experimentally
through measuring the atmospheric $\nu_\mu$ and/or the $\nu_e$ event rates as
functions of $E_\nu$ and the oscillation path length, $L=|2R_E\cos\vartheta|$
($R_E$ being the Earth radius and $\vartheta$ the zenith angle). The effect is
strongest for $\sqrt2G_FN_e=\Delta_{13}\cos(2\theta_{13})$, where according to
eqs.~(\ref{eq:theeff}) and~(\ref{eq:deleff}) $\sin^2(2\theta_{13}^\text{eff})=1$. 
This condition is met for $E_\nu\approx30\,\text{GeV}/\rho[\text{g\,cm}^{-3}]$,
where $\rho$ is the matter mass density. Typical Earth densities are between
$3\,\text{g\,cm}^{-3}$ for the crust and $13\,\text{g\,cm}^{-3}$ for the inner
core (see Fig.~\ref{fig:prem}), implying that the relevant neutrino energy
regime is a few to about $20\,\text{GeV}$.

Since experiments with sufficient target mass to measure precisely the
atmospheric neutrino flux in this energy range are too large to be magnetised,
the charge of the lepton $\ell$ produced in the relevant reactions, $\nuan_\ell
N\to\ell^\pm X$ ($N$ being the target nucleon and $X$ the hadron final-state
system), cannot be measured and hence a distinction between neutrinos and
antineutrinos is impossible. Nevertheless, a measurable net effect remains due
to the fact that the $\nu N$ and $\nubar N$ cross sections differ significantly
in the relevant energy regime, $\sigma(\nu N)\approx2\sigma(\nubar N)$.

\begin{figure}[hbt]
  \includegraphics[width=0.47\textwidth]{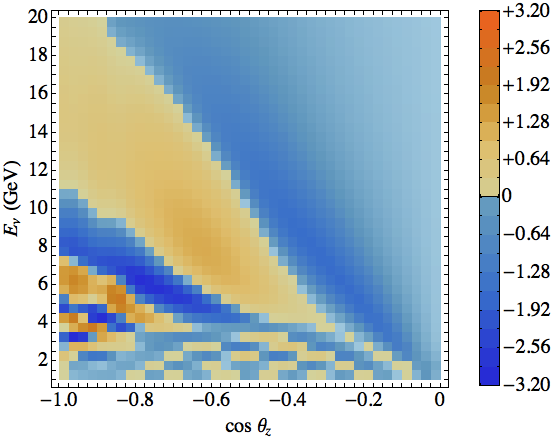}
  \hfill
  \includegraphics[width=0.47\textwidth]{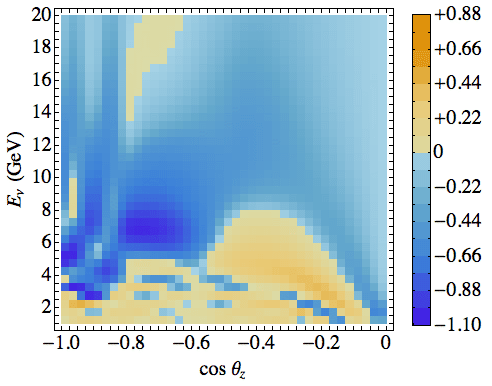}
  \caption{Pseudo-significances $(N^\text{NH}-N^\text{IH})/\sqrt{N^\text{NH}}$
           per $(E_\nu,\vartheta)$ bin, for $\nu_\mu$ (left) and $\nu_e$ events
           (right), after one year of data taking and an assumed effective
           volume increasing from 2 to 20\,Mton for neutrino energies from 2 to
           20\,GeV. Perfect experimental resolution and efficiency was assumed. 
           Plots taken from \protect\cite{ARS-2013}.}
  \label{fig:ARS}
\end{figure}

The measurement described above has e.g.\ been suggested in \cite{ARS-2013}
where the expected differences between event numbers in the NH and IH scenarios
have been calculated in bins of $E_\nu$ and $\vartheta$. The resulting
``pseudo-significances'', $(N^\text{NH}-N^\text{IH})/\sqrt{N^\text{NH}}$, are
shown in Fig.~\ref{fig:ARS} for $\ell=\mu$ (left) and $\ell=e$ (right). Even
though these plots do not include any experimental smearing and assume an
unrealistically optimistic effective detector volume, it becomes obvious that a
significant measurement might be possible also under more realistic conditions. 
Note that some of the patterns are caused by the inhomogeneous Earth density
profile (parametric resonances).

\section{The ORCA feasibility study}
\label{sec:orc}

ORCA stands for {\it Oscillation research with Cosmics in the Abyss} and denotes
the approach to perform the measurements discussed in Section~\ref{sec:atm} with
a deep-sea neutrino telescope in the Mediterranean Sea, using the technology
developed for the KM3NeT project \cite{KM3NeT-here,KM3NeT-web}. Neutrino
telescopes \cite{Katz+Spiering-2012} consist of large 3-dimensional arrays of
photo-sensors in transparent environments that register the Cherenkov light of
charged secondary particles emerging from neutrino reactions. From the arrival
time of the Cherenkov photons (nanosecond precision) and the positions of the
sensors (uncertainty about $10\,\text{cm}$), the direction and energy of the
incoming neutrino can be reconstructed.

\begin{figure}[thb]
  \sidecaption
  \includegraphics[width=0.36\textwidth]{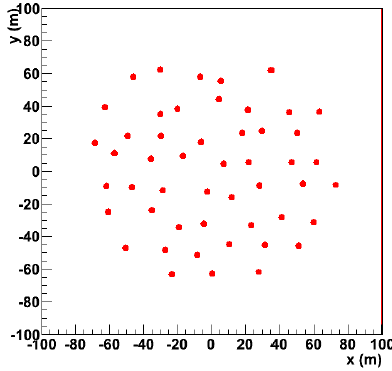}
  \hspace*{3.mm}
  \includegraphics[width=0.08\textwidth]{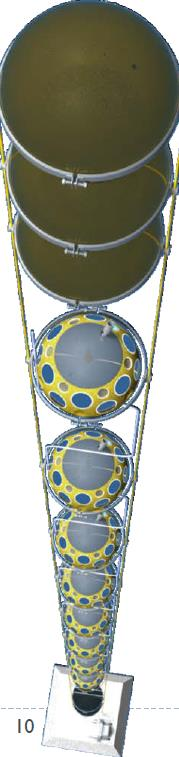}
  \caption{Footprint of the 50 detector strings assumed in the ORCA feasibility 
           study (left) and artist's view of one string (right). Note that the 
           string drawing is not to scale.}
  \label{fig:detector}
\end{figure}

A detailed feasibility study addressing the prospects of measuring the neutrino
mass hierarchy with a deep-sea neutrino telescope is being prepared by the
KM3NeT collaboration\footnote{A similar study is also pursued in the context of
the IceCube neutrino telescope in the deep ice of the South Pole: {\it Precision
IceCube Next-Generation Upgrade, PINGU} \cite{PINGU-LoI,PINGU-here}.}. For this
study, an example detector configuration was chosen with 50~strings, carrying 20
digital optical modules (DOMs) each (see Fig.~\ref{fig:detector}); each DOM is
equipped with 31 3-inch photomultipliers. The distance between neighbouring
strings is $20\,\text{m}$, the vertical distance between adjacent DOMs is
$6\,\text{m}$. The instrumented volume is $1.75\times10^6\,\text{m}^3$,
corresponding to about $1.8\,\text{Mton}$ of sea water. The detector is assumed
to be installed at a water depth of 3.5\,km. Apart from the geometrical
configuration (and the corresponding cable lengths), the detector design has
been adopted ``as is'' from KM3NeT. The technical feasibility of the selected
configuration appears likely but will have to be verified by engineering studies
before embarking on a possible future proposal.

The following performance issues are addressed in detailed simulation studies:
\begin{itemize}
\item
What are the trigger and event selection efficiencies?
\item
How can neutrino events in the relevant energy range be reconstructed and
which resolutions in $E_\nu$ and $\vartheta$ can be achieved?
\item
How can the backgrounds be controlled?
\item
How and how efficiently can the different event classes (in particular $\nu_\mu$
and $\nu_e$) be separated?
\item
What are the systematic effects and how can they be controlled?
\end{itemize}
Answers to these question are meanwhile known, even though some are still
partial or preliminary. A consistent picture emerges, indicating that a
measurement could be possible but will require an exposure (effective volume
times running time) of roughly $20\,\text{Mton}\cdot\text{year}$ for a
significance of $3\text{--}5\sigma$. A selection of findings leading to this
conclusion is presented in the following.

\looseness=-1
We start by discussing reactions of the type $\nu_\mu N\to\mu X$ (``muon
channel'') with up-going neutrino, i.e.\ zenith angle $\vartheta>90^\circ$. As
in KM3NeT, all photomultiplier signals above a noise threshold (typically
$0.3\,$photo-electrons) are sent to shore and processed by an online event
filter there. We find that the filter conditions can be adjusted such that the
random background from potassium-40 decays is almost completely suppressed and
the event selection efficiency exceeds 50\% (90\%) for neutrino energies above
$3\,\text{GeV}$ ($6\,\text{GeV}$). This result is not surprising as the muon
tracks at the relevant neutrino energies are rather short and thus the
coincidence time windows can be kept small.

\begin{figure}[hbt]
  \includegraphics[width=0.47\textwidth]{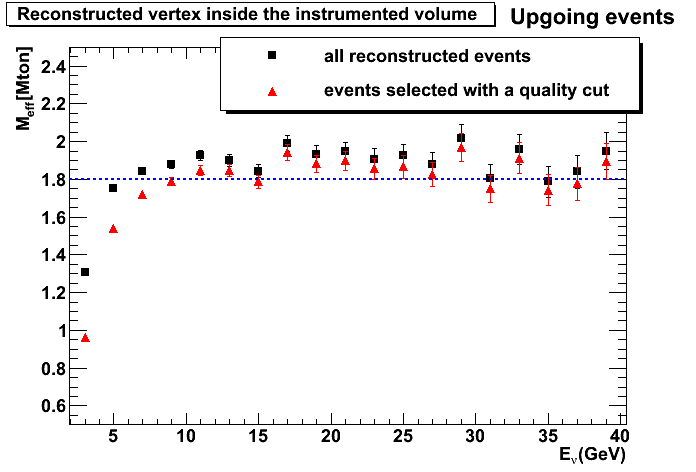}
  \hfill
  \includegraphics[width=0.47\textwidth]{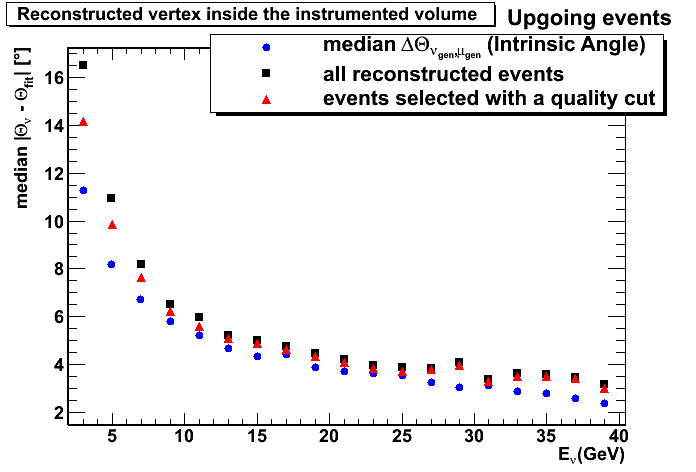}
  \caption{Left: Effective volume of the ORCA detector as a function of $E_\nu$,
           for all reconstructed and for the selected events. The instrumented 
           volume is indicated by the dashed line. The volume is given in
           units of mass, using the density of sea water. Right: Average angular
           mismatch between true neutrino and reconstructed muon direction as a 
           function of $E_\nu$, for the same two event classes. The blue symbols
           show the mismatch between the true $\nu$ and $\mu$ directions (``intrinsic
           angle'').}
  \label{fig:murec}
\end{figure}

The event reconstruction mostly uses the information of the muon track, which
has a length of about $5\,\text{m}$ per GeV of muon energy. The event
reconstruction yields the muon track length and direction, the vertex position
and a quality parameter, $\Lambda$; currently, no attempt is made to reconstruct
the hadronic shower. Well-reconstructed events are selected by cutting on
$\Lambda$ and requiring the reconstructed vertex to be inside the instrumented
volume. For the resulting event sample, the vertex is reconstructed with an
average deviation from the true value of less than 5\,m (2.5\,m) for neutrino
energies above 3\,GeV (15\,GeV). In Fig.~\ref{fig:murec} the effective volume
and the direction reconstruction precision are shown as functions of the
neutrino energy. The event selection is almost fully efficient for
$E_\nu>5\,\text{GeV}$, and the mismatch between reconstructed muon and true
neutrino direction is completely dominated by the intrinsic contribution, except
at lowest energies.

\begin{figure}[hbt]
  \sidecaption
  \includegraphics[width=0.45\textwidth]{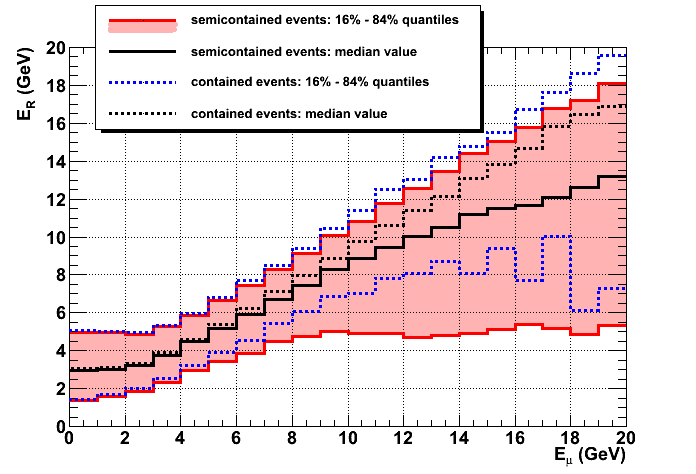}
  \caption{Muon energy reconstructed from track length, $E_R$, versus true muon
           energy. The solid black line (pink area) show the averages and the 
           $1\,\sigma$ quantiles for semi-contained events. The black (blue) 
           dashed lines indicate the same quantities for contained events.} 
 \label{fig:muone}
\end{figure}

The energy estimate is currently based on the reconstructed muon track length. 
This approach is problematic in cases where the muon leaves the detector volume
(``semi-contained events''), or in events in which a large fraction of $E_\nu$
goes into the hadronic system $X$ (high $y=(E_\nu-E_\mu)/E_\nu$). The
reconstructed muon energy, $E_\mu^\text{R}$, is shown in Fig.~\ref{fig:muone} as
a function of the true $E_\mu$, for contained and for semi-contained events. The
neutrino energy is estimated by multiplying $E_\mu^\text{R}$ with a correction
function $f(E_\mu^\text{R})$ derived from Monte Carlo simulation. The resulting
reconstructed neutrino energy, $E_\nu^\text{R}$, has a median offset from
$E_\nu$ of about 1\,GeV at $E_\nu\lesssim6\,\text{GeV}$, rising to 3\,GeV
(10\,GeV) for $E_\nu=10\,$GeV (20\,GeV). Note that for a Gaussian resolution
function the quoted median corresponds to $0.67\sigma$. An improvement of the
energy resolution is expected from properly taking into account the hadronic
system $X$; this is work in progress and results are not yet available. In the
following we will therefore make generic assumptions on the energy resolution.

\begin{figure}[thb]
  \begin{center}
  \includegraphics[width=0.325\textwidth]{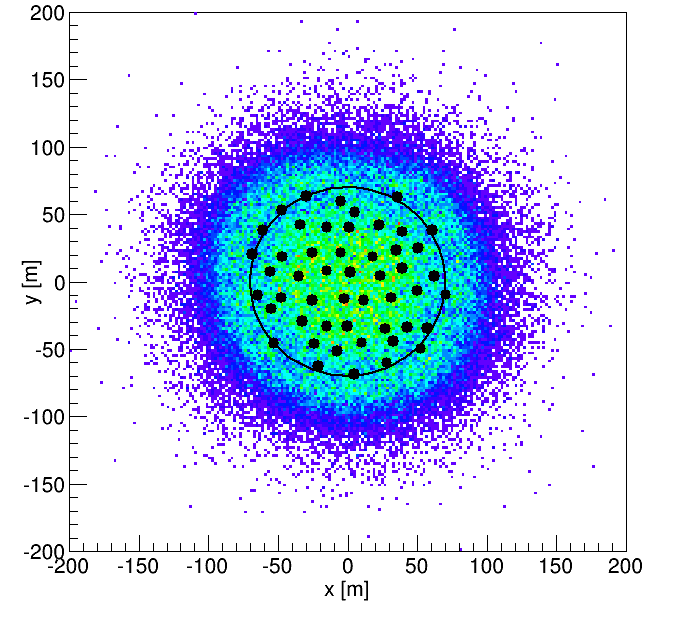}
  \hfill
  \includegraphics[width=0.325\textwidth]{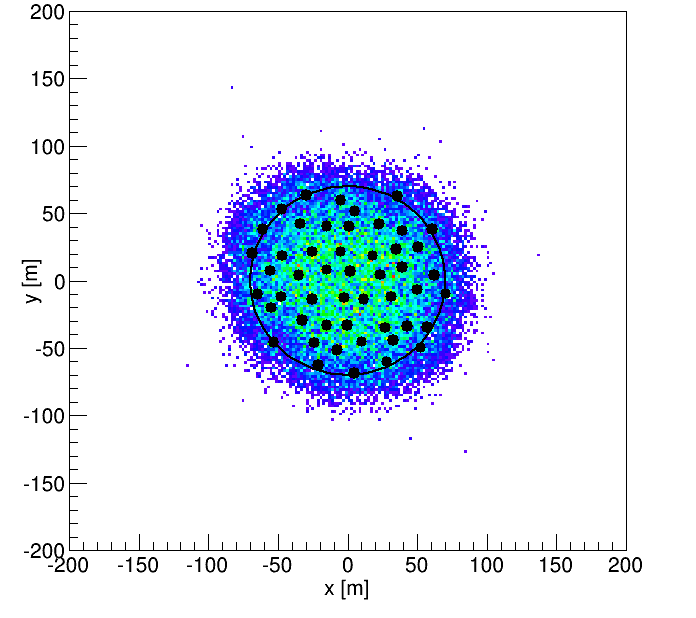}
  \hfill
  \includegraphics[width=0.325\textwidth]{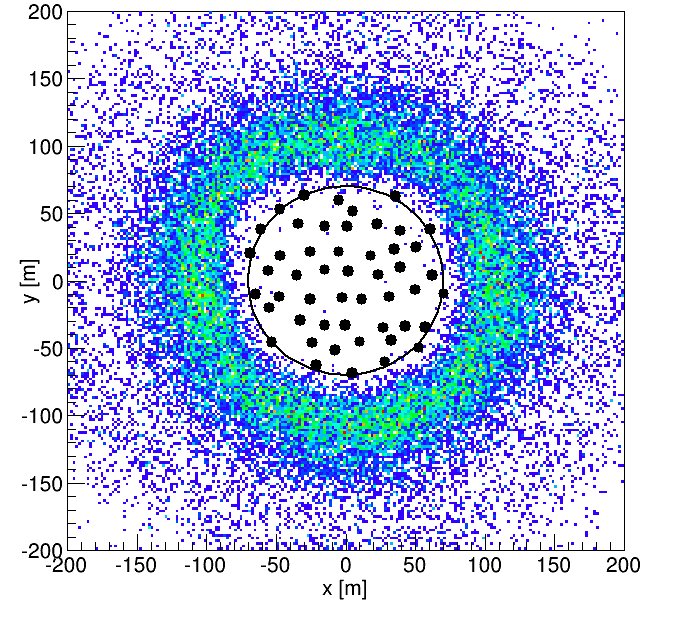}
  \end{center}
  \caption{Reconstructed vertex positions (coloured points) in the horizontal
           plane for all events reconstructed as up-going, after $\Lambda$ and
           $\beta$ cuts, for all simulated neutrino events (left), the neutrino
           events with true $E_\nu<20\,$GeV (middle) and for atmospheric muons
           (right). The detector footprint is also indicated. The black circles
           denote a possible fiducial volume cut. Note that the atmospheric
           muons are through-going and do not have a real vertex in the vicinity
           of the positions indicated.} \label{fig:atmu}
\end{figure}

The main source of background are atmospheric muons, i.e.\ those that are
generated in extended air showers in the upper hemisphere, penetrate to the
detector and are reconstructed as up-going. Even though the probability for
such misreconstruction is tiny, the resulting background can be large since
atmospheric muon events are roughly 6~orders of magnitude more frequent than
neutrino events. In absence of a surrounding detector volume to be used as veto
against atmospheric muons (as in the case of PINGU and IceCube), this background
source must be removed at the event selection/reconstruction level. It has been
found that this is well possible by applying cuts on $\Lambda$ (see above), on
the uncertainty of the reconstructed muon direction, $\beta$, as well as on the
reconstructed vertex position. Figure~\ref{fig:atmu} shows the distribution of
the reconstructed vertex position in the horizontal plane after $\Lambda$ and
$\beta$ cuts, for neutrinos and atmospheric muons. A clean separation of these
event samples is possible and the remaining background level can be adjusted to
10\% or even 1\% without a prohibitive loss of signal events.

Building on the results presented above, a simplified significance analysis was
performed by generating a large number of ``pseudo-experiments'' (PEs), i.e.\
simulated experimental measurements of event distributions in the plane of
reconstructed $E_\nu$ and $\vartheta$. These PEs cover a range of assumed
measurement durations and are generated under the following assumptions:
\begin{itemize}
\item
the neutrino interactions are generated inside the instrumented volume and at
least 15 photomultiplier hits are required;
\item
the true muon direction is used for $\vartheta$ (cf.\ Fig.~\ref{fig:murec}) and
a Gaussian uncertainty on $E_\nu$ between 10\% and 30\% is applied;
\item
no backgrounds from atmospheric muons or neutrino reactions other than in the
muon channel are considered;
\item
for each PE, NH or IH is assumed and a set of oscillation parameters
($\theta_{ij},\Delta m^2_{ij},\delta$, see Sect.~\ref{sec:osc}) is selected
according to Gaussian distributions given by the current world averages and
uncertainties of these parameters, but neglecting correlations between them.
\end{itemize}

\begin{figure}[t]
  \begin{center}
  \includegraphics[width=0.96\textwidth]{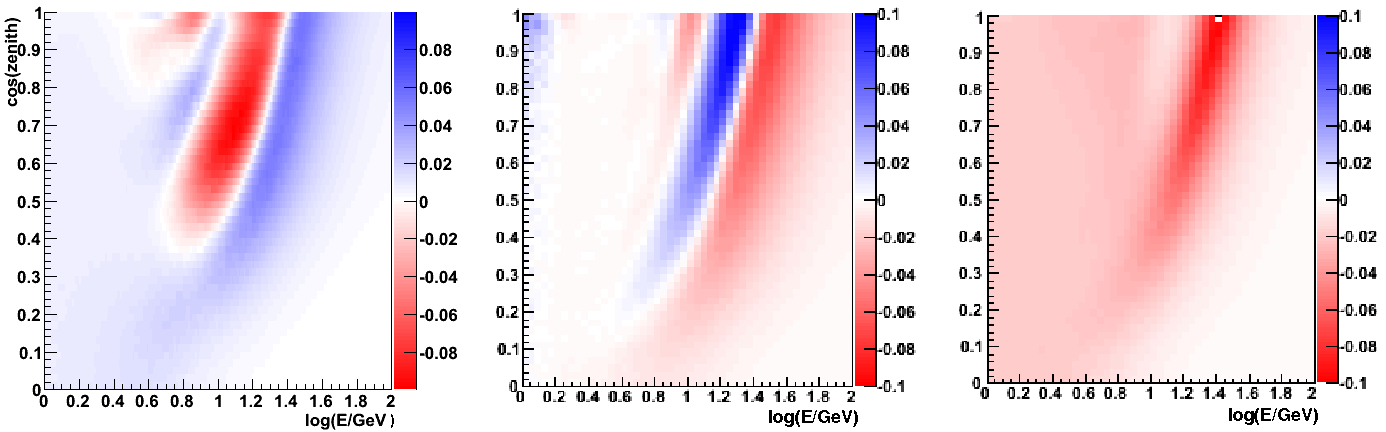}
  \end{center}
  \caption{Expected relative differences between event numbers in the NH and IH
           scenarios (left) and relative event number modifications for varying 
           $\Delta m^2_{23}$ (middle) or $\theta_{23}$ (right), respectively.}
  \label{fig:oscsys}
\end{figure}

\begin{figure}[t]
  \begin{center}
  \includegraphics[width=0.52\textwidth]{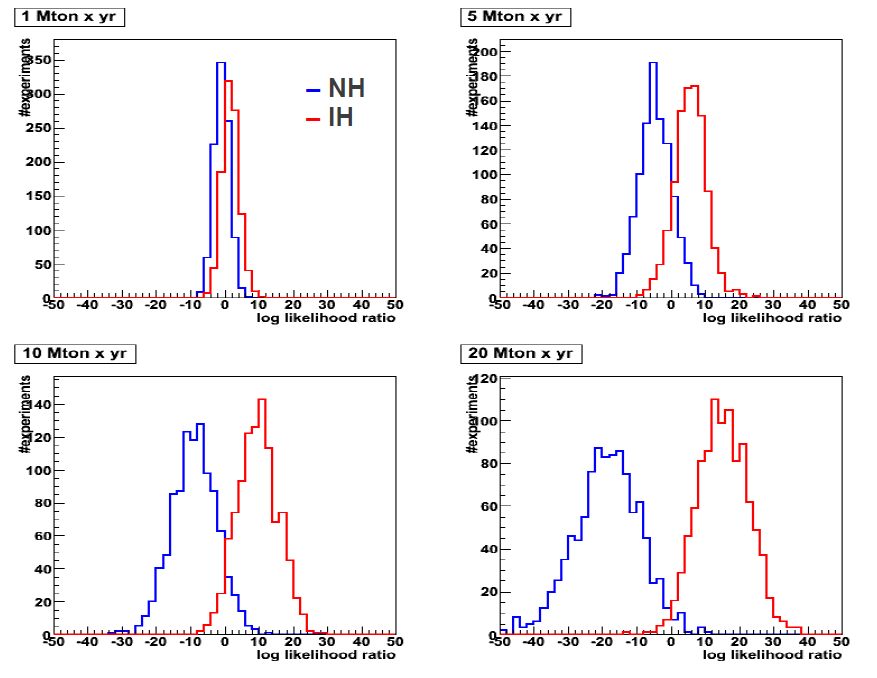}
  \hfill
  \includegraphics[width=0.4\textwidth]{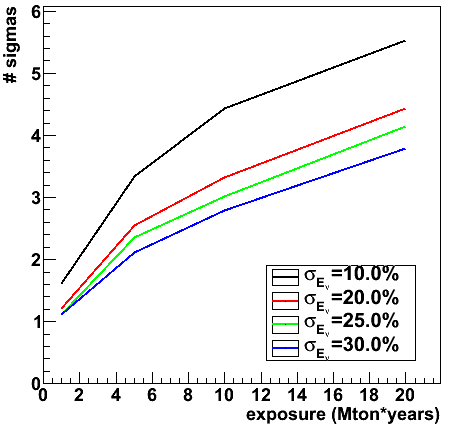}
  \end{center}
  \caption{Left: $Q$ distributions for the NH (blue) and IH (red) scenarios, for
           exposures between $1\,\text{Mton}\cdot\text{year}$ and
           $20\,\text{Mton}\cdot\text{year}$, for a Gaussian $E_\nu$ resolution
           with width 25\%. Right: Resulting significances as a function of
           exposure, for different Gaussian energy resolutions assumed.}
  \label{fig:analysis}
\end{figure}

Each PE is analysed by performing a log-likelihood fit with the oscillation
parameters as free parameters and assuming both hierarchies in turn. The maximum
likelihoods $\cal{L}$ resulting from these fits are used to calculate the
parameter
\begin{equation}
  Q=\sum_\text{bins}\log\left[\cal{L}(\text{PE data}|\text{NH})\right]-
    \sum_\text{bins}\log\left[\cal{L}(\text{PE data}|\text{IH})\right]\,,
  \label{eq:logl}
\end{equation}
which is used to quantify the separability of the NH and IH hypotheses. The main
conclusions are:
\begin{enumerate}
\item
ORCA will significantly constrain $\Delta m^2_{23}$ and $\theta_{23}$ beyond their
current precision but is rather insensitive to the other oscillation parameters;
in particular, no substantial sensitivity to the CP-violating phase $e^{i\delta}$
is found.
\item
In turn, the current experimental errors on $\Delta m^2_{23}$ and $\theta_{23}$
induce a systematic uncertainty, in particular as the patterns in the
$(E_\nu,\vartheta)$ plane caused by the variation of these parameters with in
their error bounds is similar to the NH/IH difference to be assessed (see
Fig.~\ref{fig:oscsys}). Note that this systematic uncertainty is automatically
accounted for in the statistical analysis.
\item
The distribution of the $Q$ values of all PEs is investigated to calculate the
significance of a mass hierarchy measurement as a function of the exposure. The
$Q$ distributions are shown in Fig.~\ref{fig:analysis} (left); the significance,
calculated from the distance and the widths of Gaussians fitted to the NH and
IH distributions, is presented in Fig.~\ref{fig:analysis} (right).
\end{enumerate}

Since the assumptions made are realistic, but on the optimistic side, the
resulting significances can be taken as a ``best-case scenario'' for an analysis
of the muon channel only. They are in fact compatible with the results of a
parametric study reported in \cite{Winter-2013}. We have studied in detail
sources of systematic uncertainty beyond that induced by the errors on the
oscillation parameters. It is found that those caused by the Earth density
profiles and the flux shape of the atmospheric neutrinos are small. A future
improvement in the muon channel might be possible by using the reconstructed $y$
values for $\nu/\nubar$ separation, as suggested in \cite{Ribordy-2013}.

It has recently been realised that the electron channel, i.e.\ reactions of the
type $\nu_e N\to eX$, can contribute substantially to the significance of a
neutrino mass hierarchy measurement. In fact the unsmeared NH/IH pattern
(Fig.~\ref{fig:ARS}, right) is less striking that in the muon channel, but it is
also less affected by experimental resolution effects. To make best use of this
additional analysis handle, muon and electron events have to be distinguished. 
It has been shown that in ORCA, for $E_\nu\gtrsim5\,$GeV, more than 80\% of all
muon and electron events can be classified correctly using Random Decision
Forest techniques. The resulting overall performance of ORCA is still under
investigation.

A possible future option could be a long-baseline neutrino beam targeted on
ORCA. Such a beam could e.g.\ be produced in Protvino, as discussed in
\cite{Brunner-2013}.

\section{Conclusions and Outlook}
\label{sec:con}

The deep-sea neutrino telescope technology developed in the KM3NeT project could
be used to construct a densely instrumented detector to investigate the neutrino
mass hierarchy by measuring the energy and zenith distribution of atmospheric
neutrinos. Using the muon channel alone, a 3--5$\,\sigma$ significance may be in
reach for an overall exposure of 20\,Mton$\cdot$year; including the electron
channel could significantly improve the prospects.

\acknowledgments

The author wishes to thank the organisers for a truly inspiring meeting and for 
patiently accepting this contribution long after the deadline. 

{
\bibliographystyle{./NuTel}
{\raggedright\fontsize{10.pt}{11.pt}\selectfont
\bibliography{./NuTel}}
}
\end{document}